\documentclass[journal]{IEEEtran}

\usepackage{indentfirst}
\usepackage{graphicx}
\usepackage{amsmath}
\usepackage{amssymb}
\usepackage{amsfonts}
\usepackage{mathrsfs}
\usepackage{leftidx}
\usepackage{color}
\usepackage{amsmath}
\usepackage{arydshln}
\usepackage{amsthm}
\usepackage{ragged2e}
\usepackage{cite}
\usepackage{enumerate}
\usepackage{longtable}
\usepackage{float}
\usepackage{stfloats}
\usepackage{hyperref }
\usepackage{algpseudocode}
\usepackage{algorithm}
\usepackage{subfigure}
\usepackage{tabularx}

\theoremstyle{plain}

\newtheorem{lem}{Lemma}

\theoremstyle{definition}

\usepackage{glossaries}

\usepackage[table,usenames,dvipsnames]{xcolor}
\definecolor{aa}{RGB}{175,238,238}
\definecolor{bb}{RGB}{255,255,255}

\usepackage{wasysym}

\usepackage{bm}
\usepackage{makecell}

\usepackage{siunitx}
\usepackage{multirow}   
\usepackage{array}      

\begin{document}

\title{S-MDMA: Sensitivity-Aware Model Division Multiple Access for Satellite-Ground Semantic Communication}

\author{Hui Cao, Rui Meng,~\IEEEmembership{Member,~IEEE,} Shujun Han,~\IEEEmembership{Member,~IEEE,} Song Gao,

Xiaodong Xu,~\IEEEmembership{Senior Member,~IEEE,} and Ping Zhang,~\IEEEmembership{Fellow,~IEEE}

\thanks{This work was supported in part by the National Key Research and Development Program of China under Grant 2020YFB1806905; in part by the National Natural Science Foundation of China under Grant 62501066 and under Grant U24B20131; and in part by the Beijing Natural Science Foundation under Grant L242012. \textit{(Corresponding author: Rui Meng and Shujun Han.)}}
\thanks{Hui Cao, Rui Meng, Song Gao, Xiaodong Xu, and Ping Zhang are with the State Key Laboratory of Networking and Switching Technology, Beijing University of Posts and Telecommunications, Beijing 100876, China (e-mail: caohui@bupt.edu.cn; buptmengrui@bupt.edu.cn;
wkd251292@bupt.edu.cn; xuxiaodong@bupt.edu.cn; pzhang@bupt.edu.cn).

Shujun Han is with the National Engineering Research Center for Mobile Network Technologies, Beijing University of Posts and Telecommunications, Beijing, 100876, China (e-mail: hanshujun@bupt.edu.cn).
}

}



\maketitle

\begin{abstract}
Satellite-ground semantic communication (SemCom) is expected to play a pivotal role in convergence of communication and AI (ComAI), particularly in enabling intelligent and efficient multi-user data transmission. However, the inherent bandwidth constraints and user interference in satellite-ground systems pose significant challenges to semantic fidelity and transmission robustness. To address these issues, we propose a sensitivity-aware model division multiple access (S-MDMA) framework tailored for bandwidth-limited multi-user scenarios. The proposed framework first performs semantic extraction and merging based on the MDMA architecture to consolidate redundant information. To further improve transmission efficiency, a semantic sensitivity sorting algorithm is presented, which can selectively retain key semantic features. In addition, to mitigate inter-user interference, the framework incorporates orthogonal embedding of semantic features and introduces a multi-user reconstruction loss function to guide joint optimization. Experimental results on open-source datasets demonstrate that S-MDMA consistently outperforms existing methods, achieving robust and high-fidelity reconstruction across diverse signal-to-noise ratio (SNR) conditions and user configurations.
\end{abstract}


\section{Introduction}
\label{section1}


\IEEEPARstart{W}{ith} 
the rapid deployment of low Earth orbit (LEO) satellite constellations and the integration of terrestrial and non-terrestrial networks (NTNs), satellite-ground networks are emerging as a key infrastructure for global connectivity in 6G and beyond \cite{cheng20226g,liu2018space}. These networks enable ubiquitous coverage, long-distance communication, and resilience against terrestrial failures, making them indispensable for future intelligent services such as autonomous navigation, global sensing, and emergency response \cite{xu20246g}.

To fully exploit the potential of satellite-ground networks, the convergence of communication and AI (ComAI) \cite{zhang2025comai} has been proposed. Its vision is to build communication systems that are not only data-driven but also possess semantic awareness, context adaptability, and task-oriented capabilities \cite{zhang2025comai,xu2023latent}. ComAI aims to bridge the gap between raw data transmission and intelligent decision-making by directly embedding semantic understanding into the communication process \cite{ahmed2025semantic,wang2025image,gao2025semstediff}. In this context, semantic communication (SemCom) has been recognized as a foundational technology for realizing ComAI, as it focuses on transmitting the meaning of information rather than its exact bit representation \cite{shannon1948mathematical,meng2025semantic,lu2025important,liang2025semantic,meng2025survey,xu2025semantic}.

SemCom in satellite-ground networks presents unique opportunities and challenges. On the one hand, it enables efficient transmission of critical information under bandwidth-constrained and high-latency conditions \cite{nguyen2025semantic}. On the other hand, the dynamic topology of satellite systems, various types of terminals, and limited on-board processing capabilities require lightweight, robust, and adaptive semantic codec schemes \cite{wang2025high,bian2025multi}. Recent studies have explored semantic-aware satellite relaying \cite{gao2024semantic}, semantic importance-based transmission \cite{cao2025importance}, semantic feature extraction under orbital constraints \cite{daroch2023extraction}, and joint semantic-channel optimization for satellite links \cite{hassan2024semantic}.

To support multi-user communication in satellite-ground networks, the development of effective multiple access technologies is essential \cite{shen2026semantic}. Unlike conventional bit-level multiple access methods, semantic multiple access must account for the structural characteristics, semantic importance, and recoverability of transmitted information. This paradigm shift introduces novel challenges in resource allocation and interference management \cite{zhang2023model}. Consequently, a growing body of research \cite{wu2024multi,song2025multiuser,lu2025multi} has focused on designing innovative semantic multiple access schemes to effectively address these emerging requirements.

In addition to the aforementioned semantic multiple access schemes, a new method called model division multiple access (MDMA) has also become a new research hotpot in recent years \cite{zhang2023model}. MDMA leverages AI to construct a model information space that captures both signal source and channel characteristics from a semantic perspective. By operating within a high-level, multi-dimensional semantic space, MDMA enables the extraction of both shared and personalized semantic features, thereby enhancing the efficiency and adaptability of multi-user communication. 

However, existing multiple access schemes face significant challenges when applied to satellite-ground SemCom. These include limited orthogonality in high-dimensional semantic spaces \cite{emilio20216g}, difficulty in separating overlapping semantic features under noisy conditions \cite{bourtsoulatze2019deep}, and lack of adaptive mechanisms for bandwidth-constrained semantic prioritization \cite{zhu2025task}. Moreover, most current designs assume static or terrestrial channels, which are not directly transferable to dynamic satellite-ground links with intermittent coverage, and variable signal-to-noise ratio (SNR) \cite{kodheli2021satellite}. These limitations highlight the need for novel semantic multiple access frameworks tailored to the unique characteristics of satellite-ground networks.

To address the aforementioned challenges, we propose a sensitivity-aware MDMA (S-MDMA) framework. This framework enables multi-user semantic transmission based on importance sorting under satellite-ground conditions. Through feature decomposition, sensitivity-based semantic sorting, and orthogonal multi-user transmission, the framework achieves end-to-end optimization from semantic extraction to channel adaptation. This framework not only enhances semantic recovery capabilities in low SNR environments, but also significantly reduces redundant feature transmission, while providing stronger adaptability to diverse tasks and cross-scenario generalization.

Compared with existing SemCom approaches that depend on computationally intensive gradient-based optimization such as \cite{lv2025end, pezone2025semantic}, the proposed S-MDMA framework achieves substantial complexity reduction through a modular design and lightweight sensitivity evaluation. In particular, the semantic selection module utilizes a reconstruction‑sensitivity sorting function to identify task‑relevant features, thereby eliminating the need for entropy modeling or iterative optimization procedures. Moreover, the orthogonal embedding mechanism projects shared and differential semantic components into distinct subspaces, enabling interference‑free multi‑user access and facilitating efficient bandwidth pruning. To further enhance system robustness, a carefully designed loss function ensures that users experience comparable transmission quality across varying SNR conditions, thereby promoting fairness in multi-user scenarios. Finally, the framework supports end‑to‑end inference without frequent model updates or retraining, making it especially suitable for deployment on resource‑constrained satellite platforms.
The main contributions of this paper are as follows.

\begin{itemize}
\item  We develop a S-MDMA scheme tailored for multi-user satellite-ground communication characterized by limited bandwidth and fluctuating SNR conditions. The proposed scheme decomposes semantic features into shared and differential components, ranks them based on reconstruction sensitivity, and selectively transmits the most informative features. This enables content-aware and importance-driven transmission, significantly improving semantic fidelity under dynamic channel conditions.
\item We propose a feature sensitivity sorting mechanism to quantify the contribution of each semantic feature to the reconstruction quality. This method utilizes a lightweight, sensitivity-based sorting function, avoiding gradient-based optimization and entropy modeling, thereby efficiently preserving important elements in the image while reducing computational overhead.
\item We design an orthogonal semantic embedding module that maps shared and differential semantic features into mutually orthogonal subspaces via Kronecker-based embedding. Combined with a bandwidth-pruning strategy, this module enables interference-free multi-user access and efficient utilization of constrained satellite downlink resources. 
\item Extensive simulation results on open-source datasets \cite{shao2018performance,cheng2014multi} demonstrate that the proposed S-MDMA framework outperforms typical existing SemCom schemes, deep joint source–channel coding (Deep JSCC) \cite{bourtsoulatze2019deep}, wireless image transmission transformer (WITT) \cite{yang2024swinjscc}, and MDMA \cite{liu2023semantic}. Ablation studies further validate the effectiveness of the proposed semantic sensitivity sorting and orthogonal embedding modules in improving reconstruction quality and transmission efficiency.
\end{itemize}

The rest of this paper is organized as follows. Section \ref{section2} introduces the progress of work related to ComAI. Section \ref{section3} provides the system model and overviews the proposed S-MDMA scheme. Section \ref{section4} presents the semantic similarity extraction module based on MDMA and feature sensitivity sorting mechanism. Section \ref{section5} presents the proposed semantic feature orthogonal embedding module and  error loss function for multi-user reconstruction. Section \ref{section6} gives the simulation results and analysis. Section \ref{section7} concludes this paper.

\section{Related Works}
\label{section2}

SemCom has recently attracted extensive attention as a key enabler for ComAI, especially in bandwidth‑limited and task‑oriented scenarios. In this section, we review related works from three perspectives, including SemCom, semantic multiple access, and satellite‑ground SemCom. We highlight their main ideas, strengths, and limitations, and position the proposed S‑MDMA framework within this landscape.

\subsection{SemCom}

The modern line of SemCom research is largely driven by deep learning \cite{hinton2006fast}. Xie \textit{et al.} \cite{xie2021deep} propose DeepSC, a deep learning enabled SemCom system that directly maps source data to channel symbols while preserving task‑relevant meaning rather than exact bits. Their work demonstrated that SemCom can significantly improve robustness and efficiency compared with conventional source–channel separation, especially in low‑SNR regimes. Niu \textit{et al.} \cite{niu2022paradigm} further formalize SemCom as a new paradigm for communication efficiency, focusing on task-oriented metrics and semantic fidelity instead of bit error rates. These works make the conceptual and algorithmic foundation for SemCom but mainly focus on single‑user scenarios and do not explicitly address multi‑user interference or bandwidth‑adaptive semantic selection.

Subsequent studies extend SemCom to various modalities and tasks. Xie \textit{et al.} \cite{xie2021lite} design a lite distributed SemCom system for IoT, showing that lightweight semantic encoders can be deployed on resource‑constrained devices. Weng \textit{et al.} \cite{weng2021semantic} develope SemCom for speech signals, while Liu \textit{et al.} \cite{liu2026hierarchical,liu2025lameta} explore semantic coding for text and generative SemCom with large models. Zhang \textit{et al.} \cite{zhang2022toward} complement these developments by conducting comprehensive investigations of semantic representation and compression, performing semantic transmission experiments in multiple modalities, and outlining future directions for semantic expression. These works confirm that SemCom is broadly applicable across modalities, but they typically assume relatively stable channels and do not consider the stringent bandwidth and dynamics of satellite‑ground links.

Task‑oriented SemCom has also been extensively studied \cite{xu2023task}. Liu \textit{et al.} \cite{liu2022task} propose task‑oriented SemCom for image classification, where the encoder is optimized directly for downstream task accuracy rather than reconstruction quality. Liu \textit{et al.} \cite{liu2024performance} later present a unified framework for SemCom with task‑oriented performance metrics, providing a more systematic view of semantic efficiency. These works highlight that different semantic features have different task relevance, implicitly suggesting the need for semantic importance modeling. However, they do not provide explicit mechanisms for sensitivity‑aware feature sorting or multi‑user semantic separation.

In parallel, Deep JSCC has become a core building block of many SemCom systems. Bourtsoulatze \textit{et al.} \cite{bourtsoulatze2019deep} introduce the first Deep JSCC framework for wireless image transmission, showing that end‑to‑end learned mappings can outperform traditional separate source and channel coding, especially in low‑SNR regimes. Kurka \textit{et al.} \cite{Kurka2019successive,kurlka2020deepjscc-f} extend this to Deep JSCC‑f with feedback, enabling progressive refinement and transmission of variable‑length. Xu \textit{et al.} \cite{xu2022wireless} further explore Deep JSCC for wireless image transmission in practical settings. These works demonstrate that neural encoders can learn robust representations, but they treat all features uniformly and lack explicit semantic importance modeling, which is crucial under strict bandwidth constraints.

More recent works incorporate attention and transformer architectures into JSCC. Chu \textit{et al.} \cite{chu2025agentjscc} propose attention‑enhanced JSCC for robust image transmission, leveraging attention to focus on more informative regions. Li \textit{et al.} \cite{li2023image} introduce transformer‑based JSCC for image transmission, capturing long‑range dependencies in visual semantics. These models achieve strong reconstruction performance but are often computationally heavy and less suitable for on‑board satellite deployment. 

\begin{table*}
\centering
\caption{Overview of existing SemCom research}
\label{Table.1}
\begin{tabular}{|p{3cm}|p{1.8cm}|p{5.7cm}|p{5.7cm}|}
\hline
\textbf{Research Direction}                    & \textbf{References}                                                                     & \textbf{Contributions}                                                                                                    & \textbf{Limitations}                                                            \\ \hline
\multirow{2}{*}{Foundational SemCom}           & \cite{xie2021deep}                                                                               & Propose the first deep learning–enabled SemCom system                                                                     & Focus only on single-user scenarios                                             \\ \cline{2-4} 
 & \cite{niu2022paradigm}                                                                           & Define SemCom as a new paradigm emphasizing task-oriented metrics and semantic fidelity                          & Lack mechanisms for multi-user interference mitigation and bandwidth adaptation \\ \hline
\multirow{3}{*}{Deep JSCC} & \cite{bourtsoulatze2019deep}                                                                     & Introduce the first Deep JSCC framework                                                                                   & \multirow{2}{*}{Treate all features uniformly}                                  \\ \cline{2-3}
& \cite{Kurka2019successive}, \cite{kurlka2020deepjscc-f}                                         & Extend to feedback-based DeepJSCC-f for progressive refinement                                                            &                                                                                 \\ \cline{2-4} 
& \cite{xu2022wireless}                                                                            & Explore practical wireless scenarios                                                                                      & Lack semantic importance modeling and multi-user adaptation                     \\ \hline
\multirow{2}{*}{Multimodal extensions}         & \cite{xie2021lite,weng2021semantic,liu2026hierarchical,liu2025lameta} & Extend SemCom to IoT, speech, text, and generative models                                                                 & High model complexity and assume stable channel                                 \\ \cline{2-4} 
& \cite{zhang2022toward}                                                                           & Conduct systematic investigations of semantic representation and compression across modalities                            & Limited consideration of satellite-ground scenarios and bandwidth constraints     \\ \hline
\multirow{2}{*}{Task-oriented SemCom}          & \cite{liu2022task}                                                                               & Optimize semantic encoders for downstream task accuracy (e.g., image classification)                                      & Do not address multi-user semantic separation                                   \\ \cline{2-4} 
 & \cite{liu2024performance}                                                                        & Propose a unified framework for task-oriented performance metrics                                                         & Lack sensitivity-aware feature sorting mechanisms                               \\ \hline
Attention and transformer architectures        & \cite{chu2025agentjscc}, \cite{li2023image}                                                     & Incorporate attention and Transformer architectures to capture long-range dependencies and improve reconstruction quality & Computationally heavy and limit suitability for real-time satellite deployment  \\ \hline
\multirow{2}{*}{SNR-adaptive SemCom}           & \cite{ding2021snr}                                                                               & Propose autoencoder-based SNR-adaptive JSCC                                                                               & Rely mainly on implicit adaptation                                              \\ \cline{2-4} 
 & \cite{tung2022wireless,hu2022robust,chen2024adaptive}                          & Demonstrate that semantic encoders can adjust behavior according to channel conditions                                    & Lack lightweight and explicit sensitivity-aware feature sorting                 \\ \hline

\end{tabular}
\end{table*}

SNR‑adaptive SemCom has also been investigated.  Ding \textit{et al.} \cite{ding2021snr} develop an autoencoder based on deep JSCC architecture, which is tailored to noisy channels and shows strong robustness and adaptability across users. Complementary studies \cite{tung2022wireless,hu2022robust,chen2024adaptive} confirm that semantic encoders adjust their behavior according to channel conditions, but most methods rely on implicit adaptation rather than explicit feature‑level sensitivity sorting. The overview of existing SemCom research is shown in Table \ref{Table.1}.

\subsection{Multiple Access for SemCom}

As SemCom moves from single‑user to multi‑user scenarios, multiple access becomes a critical challenge. Zhang \textit{et al.} \cite{zhang2019multiple} propose semantic beamforming for multi‑user communication, where user‑specific semantic features guide beamforming design to mitigate inter‑user interference. Their work shows that semantic information can be exploited at the physical layer, but it relies on accurate channel state information (CSI) and complex optimization, which are difficult to maintain in dynamic satellite‑ground environments.

Wu \textit{et al.} \cite{wu2024multi} develope a degraded broadcast channel (DBC)‑aware semantic fusion strategy, embedding CSI into the encoder to adaptively fuse multi‑user semantics. Song \textit{et al.} \cite{song2025multiuser} propose MU‑CSASC, a multi‑user content–style SemCom framework that separates content and style features to improve reconstruction quality across users. Lu \textit{et al.} \cite{lu2025multi} introduce intent‑aware semantic multiple access, constructing a shared intent knowledge base to support multi‑user generative communication. These works demonstrate that semantic structures (content, style, intent) can be exploited for multi‑user access, but they often suffer from scalability issues, high complexity, or reliance on generative priors. Moreover, Non‑orthogonal SemCom has also been explored. Shen \textit{et al.} \cite{shen2025qoe} improve spectral efficiency by integrating non‑orthogonal multiple access (NOMA) \cite{ding2014on}, enabling multiple users to share the same frequency band simultaneously. Zhong \textit{et al.} \cite{zhong2024exploiting} propose non‑orthogonal semantic communication (NSC), where semantic streams of multiple users are superposed in the same resource block.  While this improves spectral efficiency, semantic interference becomes a major challenge, and joint training complexity grows rapidly with the number of users. This motivates the need for more structured, orthogonal semantic representations.

Based on MDMA, Wu \textit{et al.} \cite{wu2025joint} introduce a model resource pool composed of mutually exclusive semantic models and formulate a global optimization framework to enhance the exclusiveness of mappings between different models, thereby achieving deep adversarial semantic decomposition. Further, Liang \textit{et al.} \cite{liang2024orthogonal} propose orthogonal-MDMA (O‑MDMA), which exploits the inherent noise robustness of Deep JSCC to convert “anti-channel noise capability \cite{widrow1975adaptive}” into “anti-multi-user interference capability \cite{ho2010beamforming}” , enabling multi-user access without relying on traditional physical-domain resource partitioning. In addition, Wu \textit{et al.} \cite{wu2023fusion} design a flexible image semantic fusion (FISF) module that integrates the semantic features of multiple users and employs a multilayer perceptron (MLP) to adaptively adjust feature weights, thereby accommodating heterogeneous user channel conditions. Ma \textit{et al.} \cite{ma2024semantic} propose a new discrete semantic feature division multiple access (SFDMA) scheme, which extracts the semantic information of multiple users into a discrete representation in a distinguishable semantic subspace, which enables multiple users to transmit simultaneously on the same time-frequency resource. The comparison of existing SemCom multiple access schemes is shown in Table \ref{Table.2}.

\begin{table*}
\caption{Comparison of existing SemCom multiple access schemes}
\label{Table.2}
\centering
\begin{tabular}{|p{2cm}|p{2.5cm}|p{6cm}|p{5.5cm}|}
\hline
\textbf{Modalities}          & \textbf{References}          & \textbf{Contributions}                                                                              & \textbf{Limitations}                                                                       \\ \hline
\multirow{4}{*}{Image}       & DBC-Aware \cite{wu2024multi}      & CSI‑aware semantic fusion enlarges two‑user performance region                                      & Limited to two users and lacks scalability for large networks                              \\ \cline{2-4} 
                             & MU‑CSASC \cite{song2025multiuser} & Content‑style dual‑branch fusion improves multi-user semantic reconstruction quality                & Limited scalability and high complexity for large multi-user networks                      \\ \cline{2-4} 
                             & SS-MGSC \cite{lu2025multi}        & Introduces intent‑aware semantic splitting for efficient multiuser generative communication         & Limited by high computational cost and real‑time scalability                               \\ \cline{2-4} 
                             & JDASD \cite{wu2025joint}          & Achieves stronger anti‑interference in multi‑user semantic MDMA via joint deep adversarial training & Limited scalability when the number of devices exceeds available mutually‑exclusive models \\ \hline
Word                         & NSC \cite{zhong2024exploiting}    & Introduces semantic superposition coding and decoding for efficient non‑orthogonal SemCom           & Performance limited by semantic interference and joint training complexity                 \\ \hline
\multirow{2}{*}{Cross-modal} & NOMA-QoE \cite{shen2025qoe}       & Introduces QoE‑centric resource optimization for NOMA‑based LEO–terrestrial networks                & Performance depends on accurate QoE modeling and high computational cost                   \\ \cline{2-4} 
                             & O-MDMA \cite{liang2024orthogonal} & Introduces O‑MDMA with semantic orthogonal signals for robust multiuser communication               & Scalability limited by Deep JSCC capacity and model diversity constraints                  \\ \hline
\end{tabular}
\end{table*}

\subsection{Satellite-Ground SemCom}

Satellite‑ground networks introduce additional challenges such as Doppler shifts, long propagation delays, intermittent coverage, and severe bandwidth constraints \cite{xia20246g,chen2020system,krajsic2024trends}. In order to solve these problems, there are relevant papers studying SemCom in satellite-ground scenarios. The overview of satellite-ground SemCom is shown in Table \ref{Table.3}.
In terms of physical and channel mechanism, Chen \textit{et al.} \cite{chen2024semantic} propose a method combining SemCom with orthogonal time-frequency space (OTFS) modulation to effectively alleviate the severe Doppler effect and improve transmission efficiency. Furthermore, Jiang \textit{et al.} \cite{jiang2025feature} use random noise and multi-channel mulation to improve the robustness of transmission data during the training process, so that a single training model can be adapted to remote sensing applications with different SNR and harsh environments. Yuan \textit{et al.} \cite{yuan2024deep} propose a ModNet channel guided by CSI that aligns the semantics of the image with the transmission conditions, optimizing the trade-off between the quality of the reconstruction and the utility of the network. Cao \textit{et al.} \cite{cao2025channel} improve the satellite's adaptability to different complex environments by establishing a channel codebook. 

In terms of resource and efficiency optimization, Sun \textit{et al.} \cite{sun2022deep} calculate semantic importance by analyzing the gradient of task perception results, and incorporate semantic weighting into the loss function design. Chen \textit{et al.} \cite{chen2025free} propose a scheme based on the integration of free space optical and SemCom (FSO-SC), which employs a vector quantization variational autoencoder with spatial normalization to extract key semantic features of the image while preserving complex details. Meanwhile, Zhang \textit{et al.} \cite{zhang2026elastic} propose a flexible encoding and decoding method for semantic transmission of remote sensing images. According to the SNR of the channel state, the key features are selected flexibly, and the redundant information is discarded, which significantly improves the transmission efficiency of remote sensing images. Tan \textit{et al.} \cite{tan2026adaptive} propose the adaptive residual JSCC (ARJSCC) system, which compresses remote sensing images into semantic information and residuals, achieving low-overhead transmission and high-quality reconstruction. 

In terms of architecture and network mechanism, Guo \textit{et al.} \cite{guo2025semantic} systematically analyze the SemCom architecture in 6G satellite-ground networks and propose a mission-oriented transmission mechanism to address the dynamic topology and interference issues of large-scale constellations. Nguyen \textit{et al.} \cite{nguyen2025semantic} propose a SemCom framework for NTNs, which combines denoising and gateway hopping mechanisms to improve real-time reliability in scenarios such as disaster relief. Moreover, Zhao \textit{et al.} \cite{zhao2025energy} propose probabilistic semantic transmission strategies to reduce energy consumption in space-air-ground integrated networks (SAGINs) while maintaining reliability. Huang \textit{et al.} \cite{huang2025hybrid} introduce hybrid schemes combining semantic generation with traditional bit transmission.

\begin{table*}[]
\caption{Overview of Satellite-Ground SemCom}
\centering
\begin{tabular}{|l|l|l|l|}
\hline
\textbf{Direction}                                                                          & \textbf{Sub-Direction}                                                                        & \textbf{References}    & \textbf{Contributions}                                                                \\ \hline
\multirow{4}{*}{\begin{tabular}[c]{@{}l@{}}Physical \& channel \\ mechanism\end{tabular}}       & \multirow{2}{*}{\begin{tabular}[c]{@{}l@{}}Modulation and \\ anti-interference\end{tabular}}  & \cite{chen2024semantic}   & Use OTFS modulation to mitigate the Doppler effect                                    \\ \cline{3-4} 
&                                                                                               & \cite{jiang2025feature}   & Improve robustness by employing noise and multi-channel simulation                    \\ \cline{2-4} 
& \multirow{2}{*}{Channel adaptation}                                                           & \cite{yuan2024deep}       & Propose CSI-guided ModNet alignment semantics and channels                            \\ \cline{3-4} 
&                                                                                               & \cite{cao2025channel}     & Establish a channel codebook enhances adaptability to complex environments            \\ \hline
\multirow{4}{*}{\begin{tabular}[c]{@{}l@{}}Resource \& efficiency \\ optimization\end{tabular}} & \multirow{2}{*}{\begin{tabular}[c]{@{}l@{}}Bandwidth utilization \\ improvement\end{tabular}} & \cite{sun2022deep}        & Optimize bandwidth utilization using gradient-based semantic importance calculation   \\ \cline{3-4} 
&                                                                                               & \cite{tan2026adaptive}    & Combine semantic compression and transmission enhances bandwidth flexibility \\ \cline{2-4} 
& \multirow{2}{*}{Elastic codec}                                                    & \cite{chen2025free}       & Use spatial normalization to extract essential semantic features                      \\ \cline{3-4} 
&                                                                                               & \cite{zhang2026elastic}   & Elastic encoding/decoding that flexibly selects key features based on SNR             \\ \hline
\multirow{4}{*}{\begin{tabular}[c]{@{}l@{}}Architecture \& network \\ mechanism\end{tabular}}   & \multirow{2}{*}{\begin{tabular}[c]{@{}l@{}}Mission-oriented \\ architecture\end{tabular}}     & \cite{guo2025semantic}    & Propose mission-oriented transmission mechanisms                                      \\ \cline{3-4} &                                                                                               & \cite{nguyen2025semantic} & Propose NTN framework combining gateway hopping to enhance real-time reliability      \\ \cline{2-4} 
& \multirow{2}{*}{Hybrid transmission}                                                          & \cite{zhao2025energy}                       &   Reduce power consumption in SAGINs and LEO constellations.                                                                                 \\ \cline{3-4} 
    &                                                                                               &   \cite{huang2025hybrid}                     &  Balance semantic compression with traditional bit-level reliability                                                                                     \\ \hline
\end{tabular}
\label{Table.3}
\end{table*}

\subsection{Summary}

In summary, SemCom provides powerful tools for end-to-end semantic encoding, semantic multiple access offers a promising direction for multi-user semantic separation, and satellite-ground SemCom highlights the importance of robustness and bandwidth efficiency under dynamic channel conditions. However, existing schemes do not provide stable data transmission under the constraints of satellite-ground SemCom. The S-MDMA framework proposed in this paper is designed to fill this gap, enabling bandwidth‑adaptive, interference‑resilient, and sensitivity‑aware multi‑user semantic transmission in satellite‑ground networks.

\section{System Model}
\label{section3}

In this section, we describe the the network model, channel model, and the architecture of the proposed S-MDMA.

\subsection{Network Model}
\begin{figure}[t]
\centering
\includegraphics[width=3.5in]{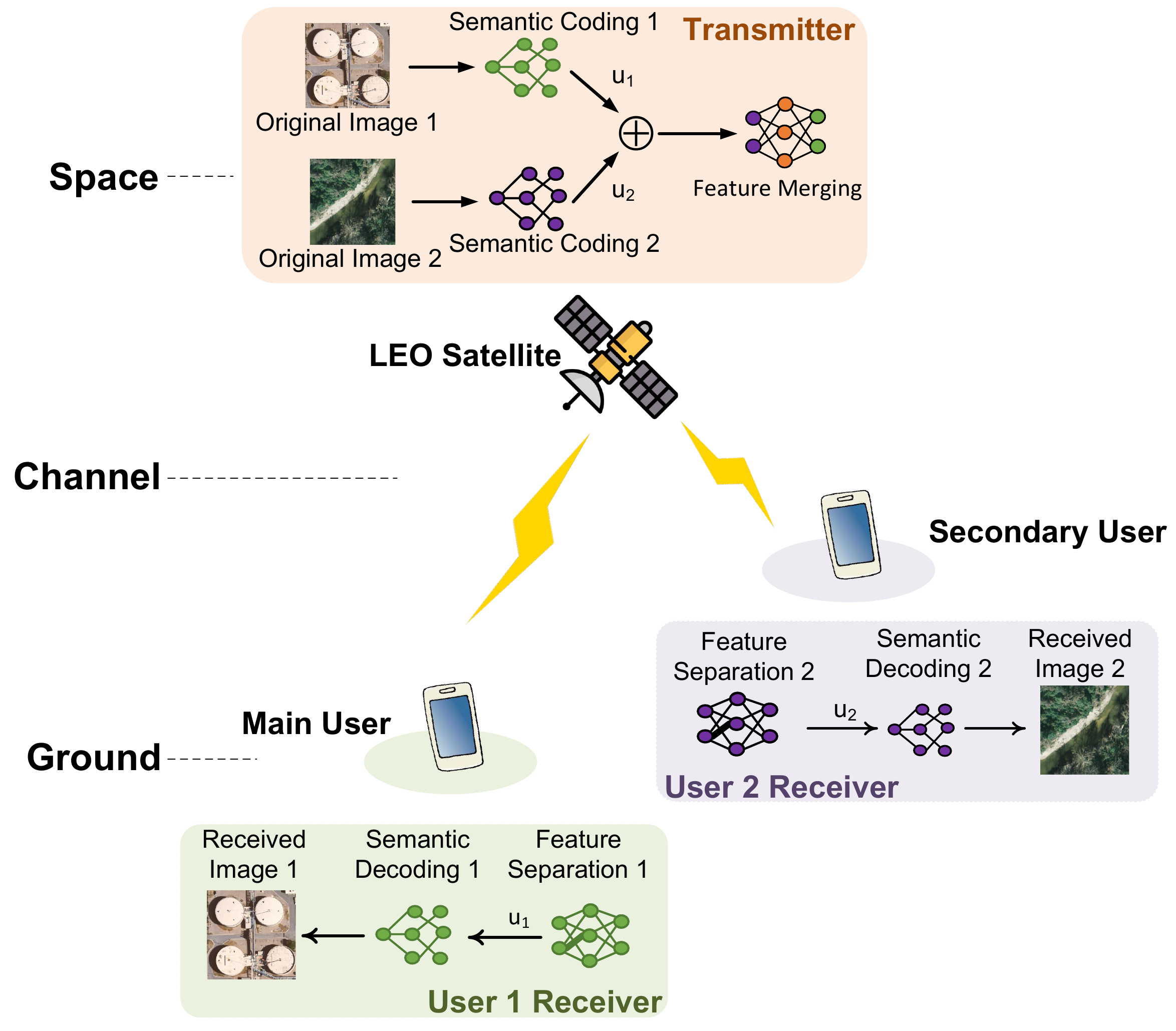}
\caption{The system model of multiuser satellite-ground SemCom, including LEO satellite, semantic transmission link, and ground terminals.}
\label{fig1}
\end{figure}

We consider a satellite-ground communication scenario in which two images are simultaneously transmitted from a satellite to two ground terminals located in different regions within its coverage area, as illustrated in Fig. \ref{fig1}. The transmission process involves several key components, which are described as follows.

\textbf{LEO satellite:} Satellites acquire raw images through local sensing or specific task data collection \cite{bui2025semantic}. Each image is then processed through semantic encoding to extract its fundamental feature representation, generating a compact semantic vector. These vectors are then organized according to semantic relevance and integrated through an orthogonal embedding mechanism, which facilitates efficient fusion while preserving the distinction between shared and differential semantic components.

\textbf{Transmission link:} The wireless link between satellites and ground terminals is inherently susceptible to environmental factors, including cloud cover and electromagnetic interference, which can significantly degrade signal quality \cite{handley2018delay}. Furthermore, due to satellites operating in different orbits and establishing connections with ground terminals at varying geographical locations, the channel characteristics are highly dynamic \cite{kodheli2021satellite}. These variations lead to fluctuations in link reliability and SNR, posing unique challenges for robust satellite-ground communication.

\textbf{Ground terminals:} Two ground users located at different positions within the satellite coverage area receive the fused downlink signal. At each user terminal, feature separation techniques are applied to extract semantic vectors related to the user's task \cite{bui2025semantic}. Subsequently, semantic decoding is performed to reconstruct the original image content, thereby achieving accurate recovery of task-relevant information within the satellite-ground communication framework. Among them, the main user needs accurate high-quality satellite images, while the secondary user has lower requirements for image quality.

\subsection{Channel Model}
Unlike traditional terrestrial networks, satellite–ground transmission channels are subject to additional impairments beyond the distance-dependent loss associated with Line-of-Sight (LoS) propagation. Specifically, the link is affected by atmospheric fading, shadowing caused by obstacles, and scattering phenomena, all of which contribute to signal degradation \cite{zhang2024space}. To accurately capture these effects, we adopt the Shadowed-Rician (SR) fading model, which has been widely applied in S-band and Ka-band satellite communication systems for data transmission \cite{adbi2003new,yuan2024deep,hong2024age}. In this paper, the satellite is assumed to employ a single transmitting antenna. Following \cite{adbi2003new}, the probability density function (PDF) of the SR channel gain, denoted as ${f_h}(r)$, is given by

\begin{equation}
\label{1}
\begin{array}{l}
{f_h}(r) = {\left( {\frac{{2{b_0}m}}{{2{b_0}m + \Omega }}} \right)^m}\frac{1}{{2{b_0}}}\exp \left( { - \frac{r}{{2{b_0}}}} \right) \\ \cdot 
{}_1{F_1}\left( {m,1,\frac{{\Omega r}}{{2{b_0}(2{b_0}m + \Omega )}}} \right).
\end{array}
\end{equation}

\noindent Here, ${b_0}$, $m$, and $\Omega$ represent the average power of the scattered component, the Nakagami‑$m$ parameter, and the LoS component, respectively. ${}_1{F_1}(a,b,c)$ denotes the confluent hypergeometric function of the first kind. In particular, for the parameter $m$, when $m$ is small, it indicates that there are more obstacles and shadows between the satellite and the ground, which may not meet the conditions for establishing the LoS link. Conversely, as 
$m$ tends to infinity, the satellite‑ground channel experiences negligible obstruction, effectively approximating a pure LoS condition. In this limiting case, the statistical properties of the SR distribution converge to those of the Rayleigh distribution, as multipath fading becomes the dominant factor in the absence of significant shadowing \cite{yuan2024deep}.

Furthermore, since the two users are situated in different regions within the same satellite coverage area, we assume that their channel parameters remain identical in this paper. However, due to spatial variations and link-specific conditions, the corresponding SNRs experienced by each user are different.

\subsection{Overview of the Proposed Sensitivity-Aware Multiple Description Multiple Access (S-MDMA) Framework}

\begin{figure*}[!t]
\centerline{\includegraphics[width=1\textwidth]{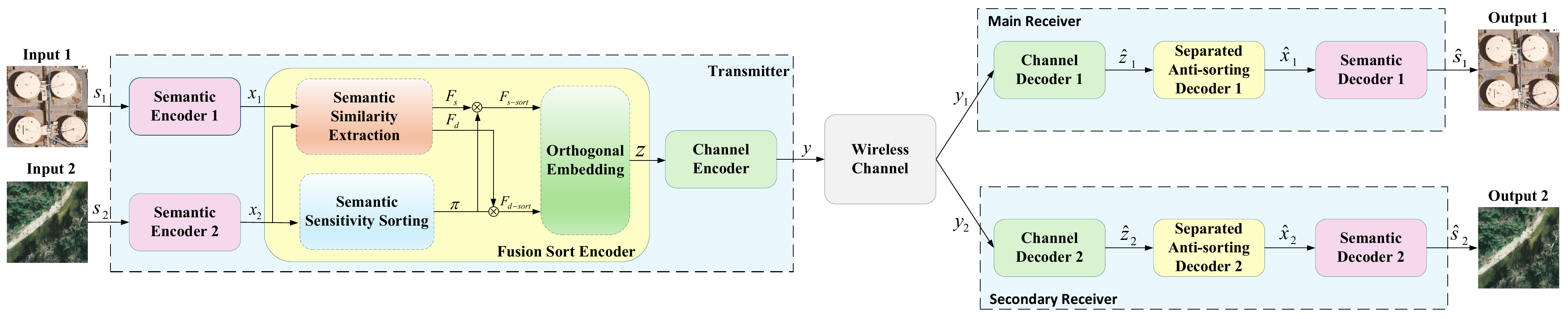}}
\caption{The architecture of the proposed S-MDMA scheme, where transmitter includes a semantic encoder, a fusion sorting encoder and a channel encoder, and receiver includes a channel decoder, a separated anti-sorting decoder, a semantic decoder.}
\label{fig2}
\end{figure*}

The framework of proposed S-MDMA scheme is shown in Fig. \ref{fig2}. The transmitter consists of a semantic encoder, a fusion sort encoder, and a channel encoder. The semantic encoder extracts high-level semantic features from the input image, while the fusion sort encoder further performs semantic similarity extraction, sensitivity sorting, and orthogonal embedding to form a structured semantic representation. The channel encoder then converts these embedded features into a transmittable signal. The receiver consists of a channel decoder, a separated anti-sorting decoder, and a semantic decoder, which are used to recover the received signal, restore the semantic feature structure, and ultimately reconstruct the semantic content of the original image, respectively.

In terms of the specific process, first, let two input images be denoted as $s_1, s_2 \in R^{H \times W \times C}$, where $s_1$ represents the main image (such as ground buildings) and $s_2$ represents the secondary image (such as forest roads). Each image is processed by an independent semantic encoder as
\begin{equation}
    x_1 = f_{\text{se}}^{(1)}(s_1),
\end{equation}
\begin{equation}
    x_2 = f_{\text{se}}^{(2)}(s_2),
\end{equation}
where $x_1, x_2 \in R^{d}$ represent the extracted semantic feature vectors, 
and $f_{\text{se}}^{(i)}$ denotes the $i$-th semantic encoder.
Then, to efficiently transmit multi-source semantic features over a shared channel, the two feature vectors are fused via a sorting-based fusion module as
\begin{equation}
    z = f_{\text{sort}}(x_1, x_2),
\end{equation}
where $f_{\text{sort}}$ denotes the fusion sort function and $z$ is the expression of $x_1$ and $x_2$ orthogonally embedded.

Next, the fused feature vector is encoded into a channel-transmittable signal in channel encoder as 
\begin{equation}
    y = f_{\text{ce}}(z).
\end{equation}
Then, the signal is transmitted to two receivers through the satellite-ground wireless channel, which is modeled as 
\begin{equation}
    y_1 = h_1(y) + n_1,
\end{equation}
\begin{equation}
   y_2 = h_2(y) + n_2,
\end{equation}
where $h_i$ denotes the channel gain and 
$n_i \sim \mathcal{N}(0, \sigma_i^2)$ is the additive Gaussian noise.

After the signal reaches the ground, the two receivers decode the received signal respectively, and obtain the estimation of the fusion feature, which is expressed as
\begin{equation}
    \hat{z}_1 = f_{\text{cd}}^{(1)}(y_1),
\end{equation}
\begin{equation}
    \hat{z}_2 = f_{\text{cd}}^{(2)}(y_2).
\end{equation}
In order to restore the original semantic features, we perform feature separation and anti-sorting operations on the decoded fusion features, which is modeled as
\begin{equation}
    \hat{x}_1 = f_{\text{unsort}}^{(1)}(\hat{z}_1), 
\end{equation}
\begin{equation}
    \hat{x}_2 = f_{\text{unsort}}^{(2)}(\hat{z}_2).
\end{equation}

Finally, restored semantic features are input into the semantic decoder respectively to obtain the final reconstructed image as
\begin{equation}
    \hat{s}_1 = f_{\text{sd}}^{(1)}(\hat{x}_1), 
\end{equation}
\begin{equation}
    \hat{s}_2 = f_{\text{sd}}^{(2)}(\hat{x}_2).
\end{equation}

\section{Proposed Semantic Similarity Fusion and Sort}
\label{section4}

In this section, we present the semantic similarity fusion and sort algorithms. First, we propose a similarity semantic feature fusion method based on MDMA. Then, we propose a sorting algorithm based on semantic reconstruction sensitivity. 

\subsection{Similarity Semantic Feature Fusion}

\begin{figure}[t]
\centering
\includegraphics[width=3.5in]{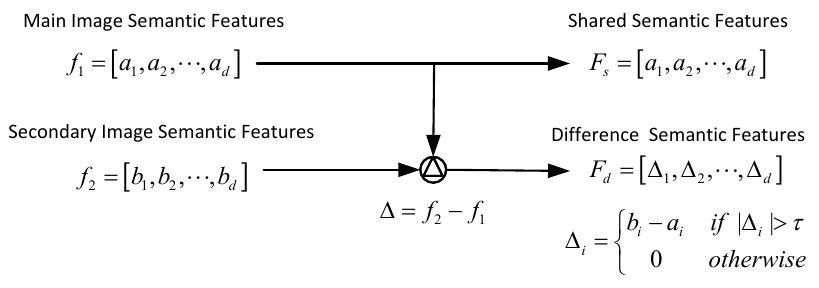}
\caption{The process of similarity semantic feature fusion.}
\label{fig3}
\end{figure}

In multi-image transmission scenarios, different input images typically share a large amount of semantic content (e.g., background, texture, or structural patterns), while also containing their own unique feature information \cite{wu2024multi}. Independently encoding and transmitting the semantic features of each image leads to redundant bandwidth consumption and reduces system robustness, especially under bandwidth-constrained or low SNR conditions \cite{peng2025robust}. To enhance semantic reuse and minimize unnecessary transmission, we perform a similarity comparison of the semantic features extracted from the two images. This enables the explicit construction of shared semantic features and differential semantic features, thereby improving transmission efficiency and reducing redundancy. This process is illustrated in Fig. \ref{fig3}.

Specifically, let the semantic feature vectors of the two images be ${f_1},{f_2} \in {R^d}$ and ${f_1} = \left[ {{a_1},{a_2}, \cdot  \cdot  \cdot ,{a_d}} \right]$, ${f_2} = \left[ {{b_1},{b_2}, \cdot  \cdot  \cdot ,{b_d}} \right]$. Since we assume that $s_1$ is the main image, we set the semantic feature $f_1$ of $s_1$ to shared semantic features $F_s$. 
Then, we calculate their difference from element to element as
\begin{equation}
    \Delta  = {f_2} - {f_1}.
\end{equation}
For each dimension $i$, if $|{\Delta _i}| \le \tau $, where $\tau$ is the similarity threshold, we consider that the dimension has a consistent semantic contribution in the two images.

For the dimensions that do not satisfy the above conditions, we believe that they contain the difference semantic information between the two images, such as the change of object shape, the difference of local texture or the difference of scene content. For each element $\Delta_i$ in $F_d$, we have 
\begin{equation}
   {\Delta _i} = \left\{ {\begin{array}{*{20}{c}}
{{b_i} - {a_i}}&{if \quad {\rm{ |}}{\Delta _i}| > \tau, }\\
0&{otherwise.}
\end{array}} \right.
\end{equation}
Through the above decomposition, $F_s$ represents the semantic content shared by the two images, while $F_d$ captures the semantic differences between them. This structured feature splitting not only reduces redundant transmission, but also enables the system to reuse public semantics in shared channels. 

\subsection{Semantic Sensitivity Sorting and Cropping Algorithm}

In the SemCom system, the contribution of semantic features of different dimensions to the final reconstruction quality is not the same \cite{wang2024feature}. Some feature dimensions are more sensitive to the structure, texture or key semantic regions of the image, while other dimensions only affect subtle or redundant information \cite{hu2023scalable,guo2025semantic2}. In order to prioritize the most critical semantic information under limited bandwidth conditions, we introduce a reconstruction-based sensitivity sorting mechanism to measure and rank the importance of each dimension of semantic features.

As mentioned above, if the semantic feature vector is $f \in R^{d}$, we perturb each dimension $i$ and observe its effect on the reconstructed image. For the $i$-th feature dimension, we construct the perturbed feature vector as 
\begin{equation}
    f^{(i)} = f + \epsilon e_i,
\end{equation}
where $e_i$ is the $i$-th standard basis vector and $\epsilon$ is the small perturbation amplitude.
Then, we input $f^{(i)}$ into the semantic decoder to obtain the reconstructed image after perturbation as 
\begin{equation}
    \hat{s}^{(i)} = f_{\text{sd}}(f^{(i)}).
\end{equation}
We use the variation of reconstruction error $\beta_i$ as the sensitivity index of this dimension, which is defined as
\begin{equation}
    \gamma_i = \mathcal{L}\big(s, \hat{s}^{(i)}\big) - \mathcal{L}\big(s, \hat{s}\big),
\end{equation}
where $\mathcal{L}(\cdot)$ is the reconstruction loss function. The greater of the sensitivity $\beta_i$ is, the more critical the dimension is to the reconstruction quality.
Finally, we sort all dimensions according to sensitivity to obtain a sorted index as 
\begin{equation}
    \pi = \text{argsort}(\beta_1, \beta_2, \ldots, \beta_d).
\end{equation}
Here, $\pi$ denotes the order of semantic dimensions from the most important to the least important.

After obtaining the reconstruction sensitivity sorting $\pi$ of all semantic feature dimensions, we further perform bandwidth-aware pruning to adapt to the semantic representation of the transmission and the available channel resources.  Since the sorted indicator set $\pi$ sorts the semantic dimensions from the most important to the least important, retaining only the top-ranked components allows the system to discard low-impact or redundant dimensions while retaining the most critical semantic information.

Let the sorted semantic feature vector be denoted as
\begin{equation}
f_{\pi} = f(\pi),
\end{equation}
where the notation $f(\pi)$ represents reordering the feature vector according to the sorting $\pi$. Given a bandwidth ratio $r \in (0,1]$, the number of preserved semantic dimensions is
\begin{equation}
K = \lfloor r d \rfloor,
\end{equation}
where $d$ is the total dimension of semantic features. 

In order to implement clipping in engineering, the system constructs a binary mask with the same dimension as the feature vector, which is 
\begin{equation}
m_i =
\begin{cases}
1, & i \le K, \\
0, & i > K.
\end{cases}
\end{equation}
Then, the sorted features are cropped by element-by-element multiplication as
\begin{equation}
    \tilde{f}_{\pi} = m \cdot f_{\pi}.
\end{equation}
This operation ensures that only the most sensitive semantic dimensions of $K$ are retained, while the remaining $d-K$ dimensions are pruned to zero. In this way, the system can maximize the retention of key semantic information under limited bandwidth, thus significantly improving transmission efficiency and robustness to reconstruction. The above flow is shown in \textbf{Algorithm}~\ref{alg1}.

\begin{algorithm}[tbp]
\renewcommand{\algorithmicrequire}{\textbf{Input:}}
\renewcommand{\algorithmicensure}{\textbf{Output:}}
\caption{Semantic Sensitivity Sorting and Cropping}
\label{alg1}
\begin{algorithmic}[1]
\Require Semantic feature vector $f \in R^d$; Semantic decoder $f_{\text{sd}}(\cdot)$; Reconstruction loss function $\mathcal{L}(\cdot, \cdot)$; Perturbation amplitude $\epsilon$; Bandwidth ratio $r \in (0, 1]$.
\Ensure Cropped semantic feature vector $f_r \in R^d$.
\State Initialize sensitivity list $\mathbf{B} = [\,]$;
\For {$i = 1$ to $d$}
     \State Construct perturbed feature vector: $f^{(i)} = f + \epsilon \cdot e_i$;
     \State Decode perturbed feature: $\hat{s}^{(i)} = f_{\text{sd}}(f^{(i)})$;
     \State Decode original feature: $\hat{s} = f_{\text{sd}}(f)$;
     \State Compute sensitivity: $B_i = \mathcal{L}(s, \hat{s}^{(i)}) - \mathcal{L}(s, \hat{s})$;
     \State Append $B_i$ to $\mathbf{B}$;
\EndFor
\State Sort indices by descending sensitivity: $\tau = \text{argsort}(\mathbf{B})$;
\State Compute number of preserved dimensions: $K = \lfloor r \cdot d \rfloor$;
\State Construct binary mask $m \in \{0,1\}^d$: $m_i = 1$ if $i \in \tau_{1:K}$, else $m_i = 0$;
\State Apply mask to sorted feature vector: $f_r = m \odot f(\tau)$;
\Return $f_r$
\end{algorithmic}
\end{algorithm}

Finally, we analyze the computational complexity of the proposed semantic sensitivity sorting algorithm. The dominant cost arises from the dimensionality $d$ of the semantic feature vector. Specifically, for each dimension, the algorithm perturbs the corresponding feature and performs a single forward pass through the semantic decoder to assess reconstruction sensitivity. This results in a total complexity of $O(d \cdot {C_{sd}})$, where ${C_{sd}}$ enotes the computational cost of one decoder inference. Once the sensitivity scores are obtained, the subsequent sorting operation incurs an additional $O(d \cdot \log d)$ overhead, which is negligible relative to the decoding cost. Consequently, the overall complexity scales linearly with the feature dimension. 

In contrast, gradient-based optimization methods typically require iterative backpropagation \cite{lv2025end}, leading to a complexity of $O(N \cdot d \cdot {C_{sd}})$, where $N$ is the number of iterations. Similarly, entropy-based approaches often involve $O(d^2)$ operations due to joint distribution estimation \cite{chalk2016relevant}. The proposed sensitivity sorting algorithm, by avoiding iterative refinement and probabilistic modeling, offers a lightweight alternative with significantly reduced computational burden. This efficiency makes it particularly well-suited for resource-constrained satellite platforms, where energy consumption, latency, and processing overhead is critical.

\section{Proposed Multi-user Data Transmission Module}
\label{section5}
In this section, we propose a data processing scheme suitable for multi-user image transmission between satellite and ground. Firstly, we introduce the semantic feature orthogonal embedding for data fusion. Then, we design a reconstruction loss function for multi-user data transmission.

\subsection{Semantic Feature Orthogonal Embedding}

In multi-user SemCom scenarios, both shared features and difference features must be superimposed and transmitted over the same channel. If these two types of feature are added directly, their high correlation in the feature space makes it difficult for the receiver to accurately disentangle the semantic components from the mixed signal, particularly under low SNR conditions \cite{zhang2025rethinking}. This often leads to semantic aliasing and the accumulation of reconstruction errors \cite{mu2022semantic}. To enable interference-free multiplexing of multiple semantic streams within shared channels, it is therefore essential to design an embedding mechanism that guaranties separability in the feature space, ensuring that distinct semantic components remain linearly independent during transmission \cite{zhang2025multimodal}. Motivated by this requirement, we introduce an orthogonal embedding mechanism that maps shared features and difference features into two strictly orthogonal subspaces. This design not only ensures mathematical recoverability but also facilitates low-complexity feature separation in practical engineering implementations.

As mentioned above, the shared feature vector sorted is $F_{s-sort}$ and the difference feature vector sorted is $F_{d-sort}$. The system constructs two orthogonal vectors of length $d$ in advance as $u_1, u_2 \in R^{d}$, which satisfies $u_1^\top u_2 = 0$.
To embed semantic features into high-dimensional orthogonal subspaces, we use Kronecker product to construct extended features as
\begin{equation}
    {F}_{s-emb} = F_{s-sort} \otimes u_1,
\end{equation}
\begin{equation}
    {F}_{d-emb} = F_{d-sort} \otimes u_2.
\end{equation}
Thus, the transmitted mixed feature is given by 
\begin{equation}
    F_{mix} = {F}_{s-emb} + {F}_{d-emb}.
\end{equation}

\begin{lem}
\label{lem}
${F}_{s-emb}$ and ${F}_{d-emb}$ are mutually orthogonal.
\end{lem}

\begin{IEEEproof}
    Using the mixed-product property of the Kronecker product \cite{charles2000ubiquitous}, the inner product between ${F}_{s-emb}$ and ${F}_{d-emb}$ can be written as
\begin{equation}
\begin{array}{l}
F_{s - emb}^ \top {F_{d - emb}}\\
 = {({F_{s - sort}} \otimes {u_1})^ \top }({F_{d - sort}} \otimes {u_2})\\
 = (F_{s - sort}^ \top {F_{d - sort}})(u_1^ \top {u_2}).
\end{array}
\end{equation}
Since $u_1$ and $u_2$ are orthogonal, we have $u_1^\top u_2 = 0$, and thus
\begin{equation}
    {F}_{s-emb}^\top {F}_{d-emb}
    = (F_{s-sort}^\top F_{d-sort}) \cdot 0 = 0.
\end{equation}
Therefore, the two embedded feature vectors are orthogonal in the high-dimensional space independent of the correlation between $F_{s-sort}$ and $F_{d-sort}$.
\end{IEEEproof}

Thus, the receiver separation process can be realized by a simple vector projection as
\begin{equation}
    \hat F_{s-sort} = (\hat F_{mix} \otimes u_1^\top),
\end{equation}
\begin{equation}
    \hat F_{d-sort} = (\hat F_{mix} \otimes u_2^\top).
\end{equation}
Power normalization is utilized to maintain the same energy as the original feature and avoid energy explosion after superposition.

\subsection{Multi-User Reconstruction Error Loss Function}

In the multi-user SemCom system, a single coded semantic representation is transmitted to multiple users who experience heterogeneous channel conditions. This setting is similar to the classical broadcast channel (BC) in information theory, where users with different SNR must simultaneously decode the information in the shared transmission \cite{lu2025self,gong2024compression}. A key challenge for such systems is to design an optimization goal to ensure balanced performance among users, prevent the model from overfitting users with high SNR, and ignore users with low SNR \cite{xie2022task}.

To address the denoising and robustness requirements inherent in this scenario, we employ a lightweight one-dimensional convolutional autoencoder as the channel codec. The channel encoder extracts high-dimensional semantic representations of the input signal through two convolutional layers with progressively increasing feature dimensions, thereby enhancing the model’s ability to capture structural information. The channel decoder adopts a symmetrical architecture to gradually reconstruct the signal, effectively suppressing noise interference and recovering the underlying clean data. Moreover, the use of a nonlinear activation function (ReLU) enhances feature discrimination, enabling the model to maintain stable reconstruction performance even under adverse low-SNR conditions. 

Traditional loss formulations, such as the arithmetic mean of user reconstruction errors \cite{tillmann2025semantic}, are often dominated by users with better channel conditions. This leads to unfair performance and reduces the robustness of the system. To solve this problem, we adopt multi-user loss based on geometric mean, naturally emphasize the worst-performing users, and encourage encoder-decoder combination learning to be robust to all users at the same time.

Assume that the transmitted semantic feature is denoted as $z$, and two users receive noisy versions of the transmitted signal under different SNR conditions. After channel decoding, the reconstructed features for user 1 and user 2 are denoted as $\hat z_1$ and $\hat z_2$. Let $L(\cdot,\cdot)$ denote the reconstruction error metric, such as mean squared error (MSE). The reconstruction losses for the two users are
\begin{equation}
    {L_1} = L({{\hat z}_1},z), 
\end{equation}
\begin{equation}
    {L_2} = L({{\hat z}_2},z).
\end{equation}
Therefore, we define the overall loss function as
\begin{equation}
{L_{all}} = \sqrt {{L_1} \cdot {L_2}}.
\end{equation}
This formulation offers several desirable properties. First, it emphasizes the worst-case user, as $L_{all}$ is dominated by the larger of $L_1$ and $L_2$, thereby discouraging the model from neglecting low SNR users. Second, the geometric mean introduces a degree of scale invariance, making the loss less sensitive to absolute magnitude differences between users. Finally, minimizing 
$L_{all}$ implicitly encourages the simultaneous reduction of both 
$L_1$ and $L_2$, leading to a more balanced and fair optimization across users.

\begin{algorithm}[tbp]
\renewcommand{\algorithmicrequire}{\textbf{Input:}}
\renewcommand{\algorithmicensure}{\textbf{Output:}}
\caption{Training Procedure for Multi-User SemCom System}
\label{alg2}
\begin{algorithmic}[1]
\Require Training dataset $\mathcal{D}$, channel encoder $f_e$, channel decoder $f_s$, SNR range $(\gamma_{min1}, \gamma_{max1})$ and $(\gamma_{min2}, \gamma_{max2})$, loss function $L$, train epoch $E_p$.
\Ensure Channel encoder $\theta_e$ and decoder $\theta_d$.
\State Add Training dataset $\mathcal{D}$;
\For {epoch = 1 : $E_p$}
     \State Generate semantic vector $z$;
     \State Put $z$ to $f_e$ generate transmit vector $y$;
     \State Randomly generate $\gamma_1$ satisfying $\gamma_1 \in (\gamma_{min1},\gamma_{max1})$;
     \State Randomly generate $\gamma_2$ satisfying $\gamma_2 \in (\gamma_{min2},\gamma_{max2})$;
     \State Calculate the output $y_1$ through wireless channel with  SNR $\gamma_1$;
     \State Calculate the output $y_2$ through wireless channel with  SNR $\gamma_2$;
     \State Put $y_1$ to $f_s$ generate semantic vector $\hat z_1$;
     \State Calculate the loss function $L_1=L(\hat z_1,z)$;
     \State Put $y_2$ to $f_s$ generate semantic vector $\hat z_2$;
     \State Calculate the loss function $L_1=L(\hat z_2,z)$;
     \State Calculate the overall loss function $L_{all}=\sqrt{L_1\cdot L_2}$;
     \State Update $\theta_e$ and $\theta_d$.
\EndFor
\end{algorithmic}
\end{algorithm}

The training procedure for optimizing model parameters is shown in \textbf{Algorithm}~\ref{alg2}. When the number of users increases from two to $M$, the overall loss function can be naturally generalized as
\begin{equation}
{L_{all}} = {\left( {\prod\limits_{i = 1}^M {{L_i}} } \right)^{1/M}},
\end{equation}
which corresponds to the geometric mean of individual user losses. This extension preserves the desirable properties discussed earlier without requiring additional architectural modifications. 

\section{Simulation Results and Analysis}
\label{section6}

In this section, we provide extensive simulation results to evaluate the performance of the proposed S-MDMA framework. We compare it with existing schemes such as Deep JSCC, WITT, and MDMA, and conduct ablation studies to validate the effectiveness of semantic sensitivity sorting and orthogonal embedding.

\subsection{Dataset and Simulation Parameters}
In this paper, we employ the dense labeling remote sensing dataset (DLRSD) \cite{shao2018performance} for training and validation. DLRSD is a densely annotated remote sensing dataset primarily designed for multi-label tasks, including image retrieval, classification, and semantic segmentation. The dataset comprises 2, 100 remote sensing images, each with a spatial resolution of 2 m and a fixed size of 256 × 256 pixels. The images are derived from GaoFen-1 and ZiYuan-3 satellite data and encompass 17 major land-cover categories, such as aircraft, bare soil, buildings, vehicles, grasslands, oceans, and ships. 
To assess the generalizability of the proposed model, we adopt the NWPU VHR-10 dataset \cite{cheng2014multi} for further validation. NWPU VHR-10 is a widely recognized and challenging dataset in the field of remote sensing image target detection. It consists of 800 high-resolution optical remote sensing images and has been extensively employed in tasks such as object detection and geospatial object recognition. 


For the S-MDMA model, we adopt the Swin Transformer architecture \cite{liu2021swin} as both the semantic encoder and decoder. The main structure of the model is shown in Table \ref{Table.4}. In addition, a lightweight SR channel is employed to enhance transmission efficiency. The primary parameter settings are summarized in Table \ref{Table.5}. All experiments are implemented within the PyTorch framework and executed on an NVIDIA A10 GPU, ensuring reproducibility and computational reliability.

\begin{table}[t]
\caption{Main Structure of the Model}
\begin{tabular}{ccc}
\hline
\multicolumn{1}{l}{}         & \textbf{Layers}                                                                                                               & \multicolumn{1}{l}{\textbf{Output size}}                           \\ \hline
\textbf{Semantic Encoder}    & \begin{tabular}[c]{@{}c@{}}PatchEmbed\\ SwinTransformerBlock x 2\\ SwinTransformerBlock x 4\\ LayerNorm\\ Linear\end{tabular} & \begin{tabular}[c]{@{}c@{}}256\\ 128\\ 256\\ 256\\ 96\end{tabular} \\ \hline
\textbf{Fusion Sort Encoder} & \begin{tabular}[c]{@{}c@{}}SwinTransformerBlock x 4\\ SwinTransformerBlock x 2\\ PatchReverseMerging\\ Linear\end{tabular}    & \begin{tabular}[c]{@{}c@{}}256\\ 128\\ 128\\ 256\end{tabular}      \\ \hline
\textbf{Channel Encoder}     & \begin{tabular}[c]{@{}c@{}}Conv1d+ReLU\\ Conv1d+ReLU\end{tabular}                                                             & \begin{tabular}[c]{@{}c@{}}64\\ 128\end{tabular}                   \\ \hline
\textbf{Channel Decoder}     & \begin{tabular}[c]{@{}c@{}}Conv1d+ReLU\\ Conv1d\end{tabular}                                                                  & \begin{tabular}[c]{@{}c@{}}64\\ 1\end{tabular}                     \\ \hline
\textbf{Semantic Decoder}    & \begin{tabular}[c]{@{}c@{}}SwinTransformerBlock x 4\\ SwinTransformerBlock x 2\\ PatchReverseMerging\\ Linear\end{tabular}    & \begin{tabular}[c]{@{}c@{}}256\\ 128\\ 128\\ 256\end{tabular}      \\ \hline
\end{tabular}
\label{Table.4}
\end{table}

\begin{table}[t]
\centering
\caption{Simulation Parameters}
\begin{tabular}{cc}
\hline
\textbf{Parameter Name}                   & \textbf{Value}             \\ \hline
Average Power of Sattered Component $b_0$ & 0.158                      \\
Nakagami-$m$                              & 19.4                       \\
LOS Component $\Omega$                    & 1.29                       \\
Batch Size                                & 8                          \\
Optimizer                                 & Adam                       \\
Base Loss Function                        & MSE                        \\
Learning Rate                             & 0.0001                     \\
Train Epoch                               & 100                        \\
User1 SNR Range                           & -10 dB - 0 dB              \\
User2 SNR Range                           & 0 dB - 10 dB               \\
Activation Function                       & ReLU                       \\
Weight Initialization                     & Xavier                     \\
Orthogonal Matrix $u_1$                   & {[}0.5, -0.5, 0.5, -0.5{]} \\
Orthogonal Matrix $u_2$                   & {[}0.5, 0.5, -0.5, -0.5{]} \\ \hline
\end{tabular}
\label{Table.5}
\end{table}

\subsection{Comparison Schemes}
\begin{itemize}
\item Deep JSCC \cite{bourtsoulatze2019deep} adopts an end-to-end deep neural network framework that directly maps image pixels to channel input symbols, effectively bypassing the conventional processes of image compression and channel coding. Typically structured as an autoencoder, it comprises a convolutional neural network (CNN)-based encoder and decoder, with a non-trainable channel simulation module.
\item WITT \cite{yang2024swinjscc} adopts Swin Transformer to extract the hierarchical semantic features of the image, and performs feature compression and mapping through the attention mechanism, thereby enhancing the modeling ability of the image structure. The encoder converts the image into an embedded vector, and then performs channel mapping through a multi-head attention module, and the decoder recovers the image from the received signal and maintains high fidelity.
\item MDMA \cite{liu2023semantic} directly extracts and transmits semantic features at the physical layer, effectively addressing the ``cliff effect" in traditional communication. By analyzing the importance of the encoded semantic vectors, semantically unimportant parts are discarded, allowing for flexible code rate adjustment without retraining the network. 
\end{itemize}

\subsection{Performance Metrics}

\subsubsection{PSNR}
Peak SNR (PSNR) is a widely used metric for evaluating the transmission quality of SemCom \cite{yuan2024deep,bourtsoulatze2019deep,yang2024swinjscc}. It quantifies the degree of distortion by measuring the pixel-level differences between the original image and the reconstructed image. 
The higher the PSNR value, the higher the similarity between the two images, the better the quality. It is calculated as
\begin{equation}
\label{28}
PSNR = 10 \cdot lo{g_{10}}\left(\frac{{ma{x^2}}}{{MSE}}\right),
\end{equation}
where $max$ represents the maximum possible value of the pixel value, usually 255 in image processing, and $MSE$ represents the mean square error.

\subsubsection{SSIM}
Structural similarity index measure (SSIM) \cite{zhou2004image} is a perceptual metric designed to approximate the characteristics of human visual perception in the evaluation of digital image quality. Unlike traditional pixel-based measures, SSIM assesses similarity by jointly considering three key components, luminance, contrast, and structural information, thereby providing an evaluation that is more closely aligned with the human visual system \cite{yang2024swinjscc,sagduyu2024joint,wu2024ccdm}. 
SSIM ranges from 0 to 1, with values approaching 1 denoting a higher degree of similarity between the reference and reconstructed images, and thus superior perceptual quality. It is calculated as

\begin{equation}
\label{29}
SSIM(x, y) = \frac{(2\mu_x\mu_y + c_1)(2\sigma_{xy} + c_2)}{(\mu_x^2 + \mu_y^2 + c_1)(\sigma_x^2 + \sigma_y^2 + c_2)},
\end{equation}

\noindent where $\mu_x$ and $\mu_y$ are the average values of $x$ and $y$, $\sigma_x^2$ and $\sigma_y^2$ are the variances of $x$ and $y$, and $\sigma_{xy}$ is the covariance of $x$ and $y$. $c_1$ and $c_2$ are constants used to maintain stability, usually set to $c_1 = (k_1L)^2$ and $c_2 = (k_2L)^2$, where $L$ is the dynamic range of pixel values, $k_1$ and $k_2$ are default values, generally take $k_1$ = 0.01 and $k_2$ = 0.03.

\subsection{Results Analysis}

\subsubsection{Ablation Experiment}

\begin{figure}
\centerline{\includegraphics[width=0.5\textwidth]{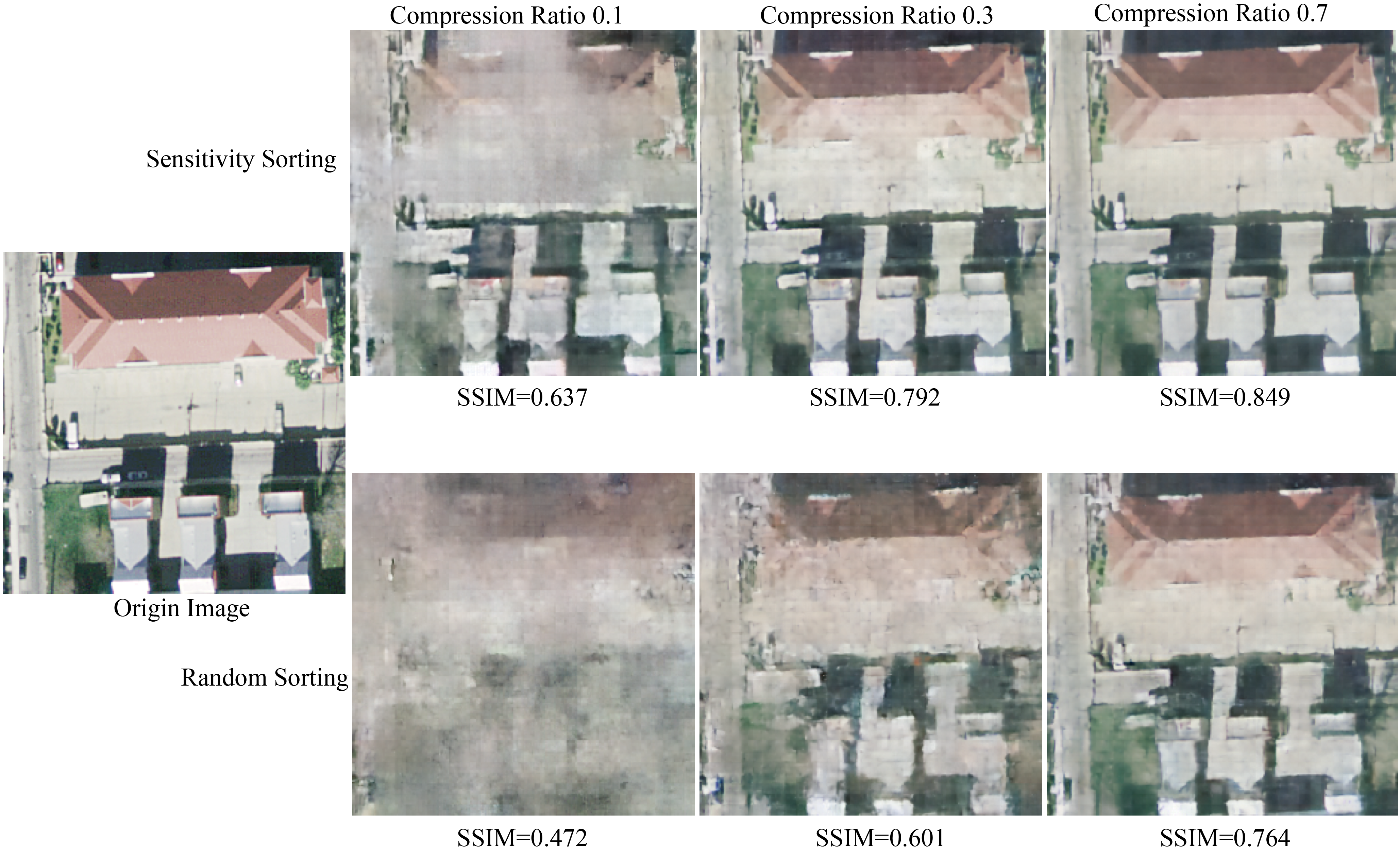}}
\caption{Image restoration comparison using different sorting schemes under different bandwidth compression ratios. By sorting semantic features by sensitivity and then performing bandwidth pruning, the system can achieve adaptive semantic compression under different bandwidth constraints, effectively preserving key structural information and significantly improving image reconstruction quality.}
\label{fig4}
\end{figure}

Fig. \ref{fig4} illustrates the comparative impact of sensitivity sorting and random sorting on the reconstruction quality of satellite images within a SemCom system across varying compression ratios. The results demonstrate that sensitivity sorting consistently achieves higher SSIM values than random sorting at all compression levels. Notably, even under low compression ratios, reconstructions obtained through sensitivity sorting retain clearly discernible image contours. This improvement arises because sensitivity-based cropping prioritizes the preservation of critical semantic structures, such as edges, contours, and spatial layouts, whereas random cropping often discards these essential features. Collectively, these findings highlight the robustness of sensitivity sorting as a bandwidth-constrained transmission strategy, ensuring superior semantic fidelity under limited resource conditions.

\begin{figure}
\centering
\includegraphics[width=3.4in]{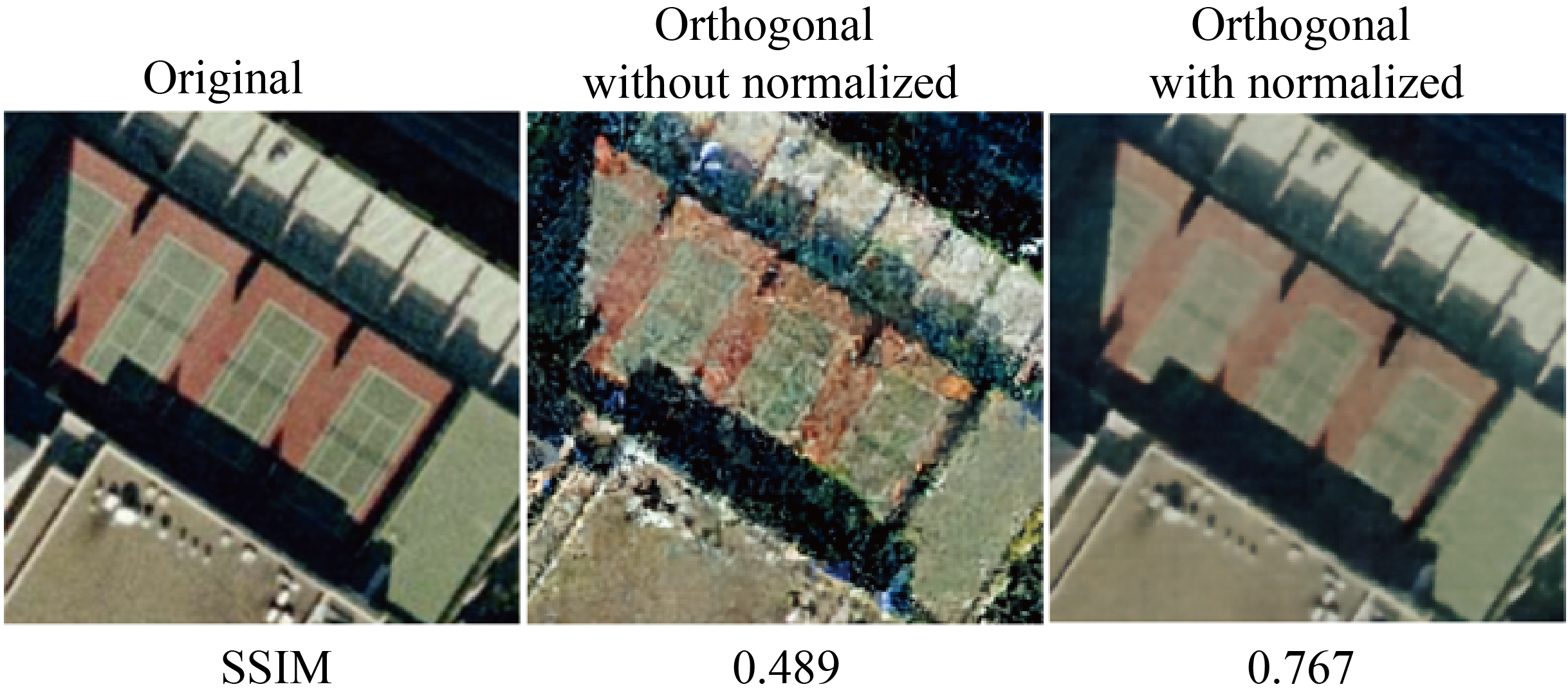}
\caption{The comparison of the reconstruction results of the same satellite image region with and without normalization. The reconstruction results obtained by normalization exhibit significantly better visual quality.}
\label{fig5}
\end{figure}

Fig. \ref{fig5} presents comparison of the reconstruction results of the same satellite image region with and without normalization. The unnormalized reconstruction exhibits pronounced degradation in both color fidelity and structural detail, characterized by color distortion and blurred edges, particularly along the tennis court boundaries and rooftop structures. In contrast, the reconstruction obtained with orthogonal normalization demonstrates markedly superior visual quality, with enhanced detail preservation and improved color consistency. This improvement can be attributed to the role of normalization in SemCom. Typically, the transmitter extracts high-dimensional semantic features of the image for compression and transmission. Without appropriate constraints, these features are susceptible to loss of discriminative power under channel perturbations or bandwidth limitations, which in turn compromises reconstruction quality. Orthogonal normalization addresses this issue by enforcing orthogonality among feature vectors, thereby enhancing their distinctiveness and stability in the feature space. As a result, the decoder is better able to recover semantic information, leading to higher-fidelity image reconstruction.

\subsubsection{Performance Comparison}
\begin{figure}[tbp]
  \centering
  \subfigure[PSNR performance comparison of different schemes.]{
  \label{fig6a}
  \includegraphics[width=0.4\textwidth]{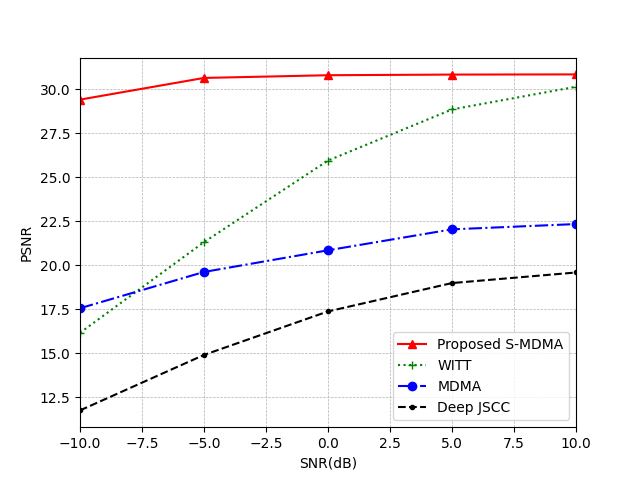}}
  \subfigure[SSIM performance comparison of different schemes.]{
  \label{fig6b}
  \includegraphics[width=0.4\textwidth]{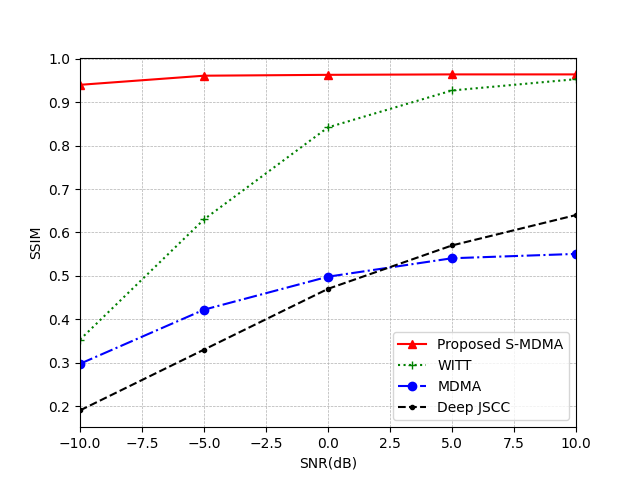}}
  \caption{The performance comparison of different schemes under varying SNR conditions.}
  \label{fig6}
\end{figure}

Fig. \ref{fig6} presents a comprehensive comparison of image transmission schemes under varying SNR conditions, evaluated using PSNR and SSIM metrics.
Fig. \ref{fig6a} illustrates the PSNR performance across the SNR range from –10 dB to 10 dB. All methods exhibit a monotonic increase in PSNR with rising SNR, reflecting improved reconstruction fidelity under cleaner channel conditions. Among them, the proposed S-MDMA scheme consistently outperforms baseline methods across all SNR values. Notably, it maintains a PSNR above 28 dB even under severely degraded conditions (e.g., –10 dB), highlighting its robustness to channel noise. This exceptional performance stems from its modulation and decoding strategy based on prioritized low SNR image reconstruction, which preserves critical image features during transmission.
Fig. \ref{fig6b} depicts the SSIM variation with respect to SNR. The proposed S-MDMA method achieves near-perfect SSIM values ($>0.95$) across all SNR levels, indicating strong preservation of structural similarity between original and reconstructed images. In contrast, competing methods show more pronounced SSIM degradation under low SNR. The stability of S-MDMA stems from its adaptive semantic model integration, which prioritizes image reconstruction under low SNR conditions. This dynamic architecture allows S-MDMA to maintain high structural fidelity while adapting to channel quality, making it well-suited for bandwidth-constrained and noisy environments.
In summary, the proposed S-MDMA architecture demonstrates superior performance in both PSNR and SSIM metrics across diverse SNR conditions. Its consistent quality preservation and noise robustness affirm its suitability for satellite-ground SemCom tasks.

\subsubsection{Adaptation Comparison}
\begin{figure}[tbp] 
  \centering
  \subfigure[PSNR performance comparison in using different datasets.]{
  \label{fig7a}
  \includegraphics[width=0.36\textwidth]{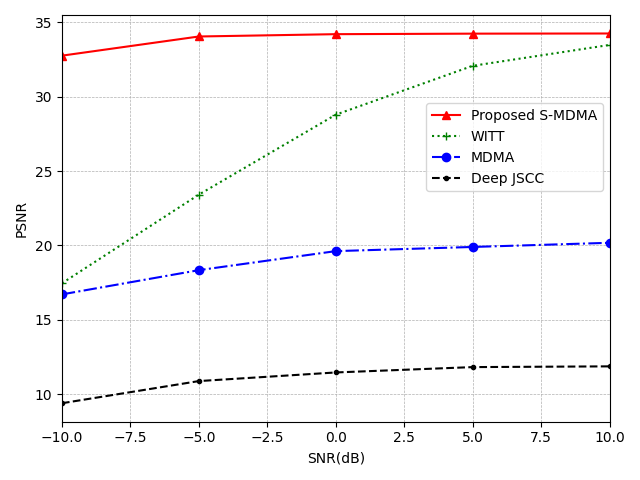}}
  \subfigure[SSIM performance comparison in using different datasets.]{
  \label{fig7b}
  \includegraphics[width=0.36\textwidth]{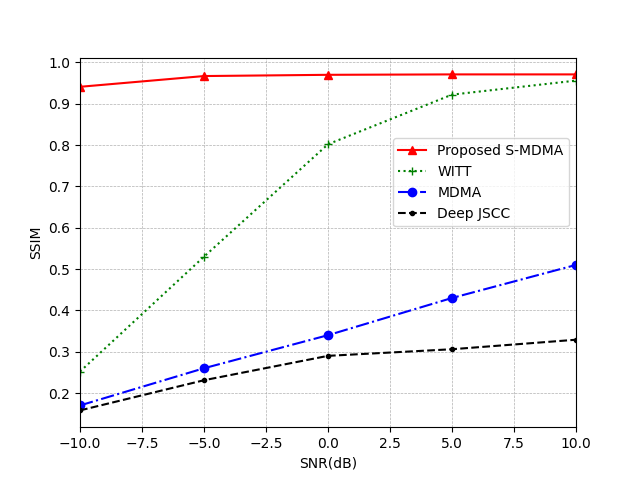}}
   \caption{The adaptation comparison of different schemes while training dataset is DLRSD and testing dataset is NWPU VHR-10.}
  \label{fig7}
\end{figure}

Fig. \ref{fig7} presents comparative results on heterogeneous datasets, where the transmission models are trained on the DLRSD dataset and evaluated on the structurally distinct NWPU VHR-10 dataset. 
Fig. \ref{fig7a} depicts the PSNR performance of four representative image transmission schemes. The proposed S-MDMA method consistently achieves the highest PSNR throughout all SNR values, underscoring its superior noise resilience and fidelity in image reconstruction. WITT ranks second, showing competitive performance particularly in high SNR regimes. MDMA demonstrates moderate effectiveness, while Deep JSCC suffers significant degradation in low SNR conditions due to its limited semantic feature extraction capability and shallow architecture.
Fig. \ref{fig7b} illustrates the SSIM curves under identical SNR settings. The proposed S-MDMA method maintains optimal SSIM values across all SNR levels, indicating a robust preservation of structural and perceptual quality despite heterogeneity of the dataset. WITT and MDMA exhibit reasonable structural consistency in higher SNR regions, whereas Deep JSCC shows poor SSIM retention under low SNR, reflecting its vulnerability to noise and limited cross-domain adaptability.
Collectively, these results validate the efficacy of the proposed S-MDMA architecture in heterogeneous remote sensing images. Unlike traditional SemCom schemes that typically require retraining or fine-tuning for specific tasks, S-MDMA enables efficient reuse of semantic representations in heterogeneous images. This property significantly reduces the need for frequent model updates, which is particularly beneficial in satellite-ground systems where on-board computational resources are limited and real-time retraining is impractical. 


\subsubsection{Effectiveness Comparison}
\begin{figure}[tbp]
  \centering
  \subfigure[PSNR performance comparison under varying compression ratios.]{
  \label{fig8a}
   \includegraphics[width=0.4\textwidth]{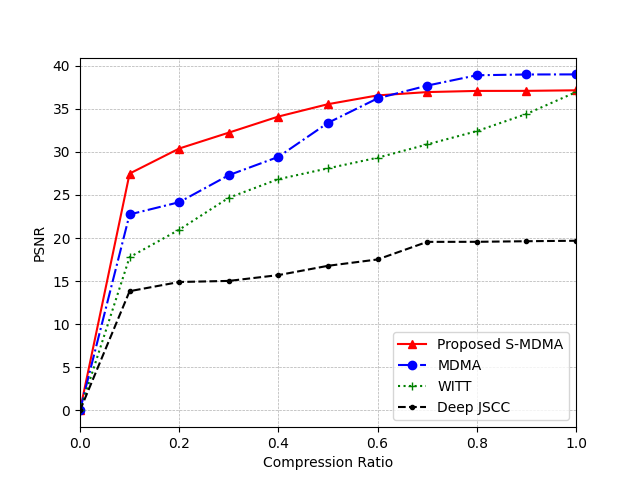}}
  \subfigure[SSIM performance comparison under varying compression ratios.]{
  \label{fig8b}
  \includegraphics[width=0.4\textwidth]{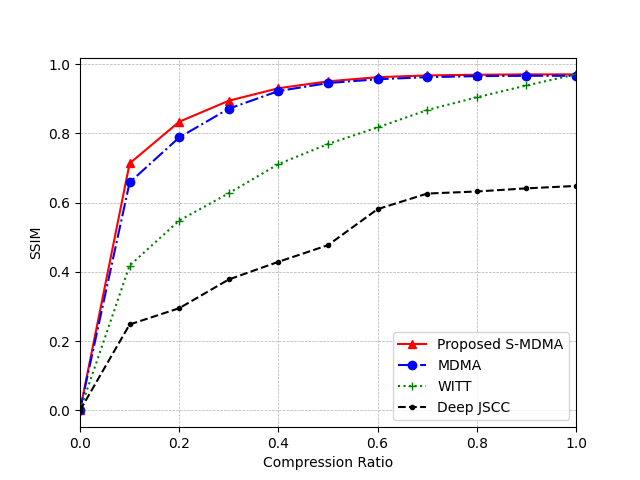}}
  \caption{The performance comparison of different schemes under varying compression ratios.}
  \label{fig8}
\end{figure}

Fig. \ref{fig8} presents comparative evaluation under varying compression ratios. An ideal channel is assumed in this experiment to highlight the impact of bandwidth constraints on image reconstruction quality.
Fig. \ref{fig8a} illustrates the PSNR trends as the compression ratio increases from 0.0 to 1.0. The PSNR of all methods increases with increasing compression ratio, reflecting the trade-off between transmission efficiency and reconstruction fidelity. The proposed S-MDMA method consistently achieves the highest PSNR across the entire compression range, maintaining values above 35 dB even at moderate compression levels (e.g., 0.5). This superior performance demonstrates its robustness in preserving pixel-level fidelity under low compression ratio, which can be attributed to its semantic sensitivity sorting strategy that ensures critical image features are transmitted with higher importance.
Fig. \ref{fig8b} shows the corresponding SSIM variations under the same compression settings. S-MDMA again demonstrates superior performance, maintaining the highest SSIM values across all compression ratios. This indicates strong preservation of structural similarity and perceptual quality even under severe bandwidth limitations. In contrast, Deep JSCC and WITT experience substantial SSIM degradation at lower compression ratios, revealing their limited capability in retaining semantic features. MDMA also benefits from semantic importance modeling and therefore performs relatively well.


\subsubsection{Visualization Result}

\begin{figure*}[!t]
\centerline{\includegraphics[width=1\textwidth]{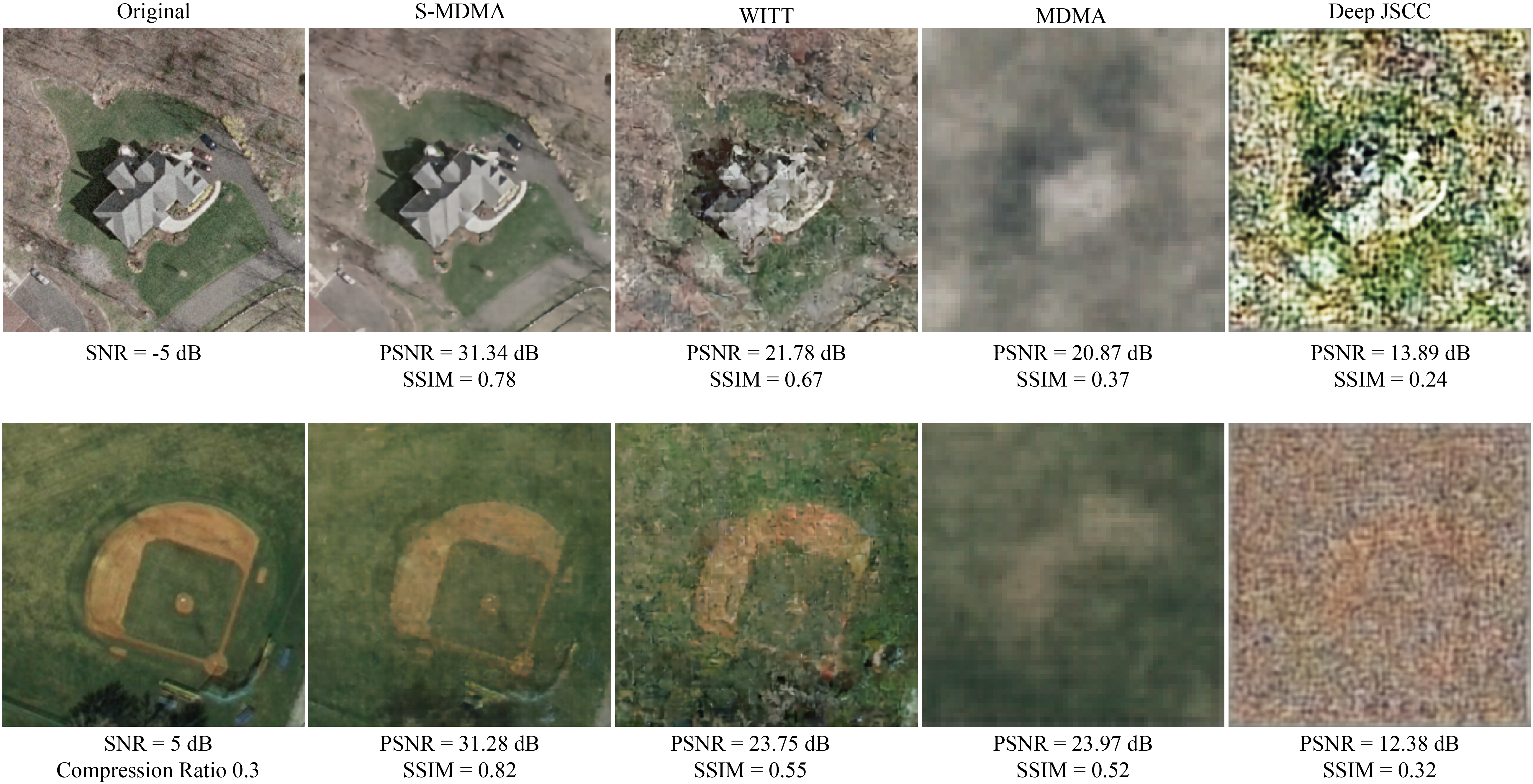}}
\caption{The visual comparison of different reconstruction schemes. The first row depicts a residential scene under a low SNR environment, while the second row illustrates a baseball field captured under a high SNR condition with a compression ratio of 0.3. Across both scenarios, the proposed S-MDMA model exhibits superior image restoration capabilities, effectively preserving structural integrity and semantic details despite varying noise levels and scene complexities.}
\label{fig9}
\end{figure*}

To further illustrate the impact of different approaches on image reconstruction performance, Fig. \ref{fig9} provides visual comparison using actual satellite remote sensing images.
The first row presents reconstruction results for a residential scene subjected to a low SNR condition of –5 dB. Under this severe noise environment, most baseline methods exhibit notable degradation in structural fidelity and semantic clarity. The Deep JSCC scheme represents the most degraded instance, reveals significant pixelation and loss of spatial coherence, rendering the house and surrounding features nearly unrecognizable. The MDMA scheme shows varying degrees of blurring and color distortion, with only coarse outlines of the building and vegetation preserved. 
In contrast, the proposed S-MDMA model maintains clear building contours and vegetation textures, demonstrating strong resilience to noise and effective semantic feature retention. This performance suggests that S-MDMA is well-suited for robust image reconstruction in low SNR transmission scenarios.
The second row presents reconstruction results for a baseball field scene under a high SNR condition of 5 dB and a compression ratio of 0.3. In this relatively favorable transmission environment, all methods exhibit improved reconstruction quality. However, notable disparities persist in the preservation of semantic details. Approaches employing random bandwidth pruning, such as MDMA, Deep JSCC, and WITT, demonstrate varying degrees of semantic degradation. Deep JSCC, in particular, suffers from geometric distortions and color inconsistencies, failing to accurately reconstruct the field’s circular layout and boundary lines. While MDMA and WITT achieve better structural recovery, they still fall short in capturing fine-grained spatial features with clarity. In contrast, the proposed S-MDMA model, which incorporates semantic sensitivity sorting algorithm, delivers markedly superior reconstruction performance. It effectively restores the geometric contours, infield configuration, and surrounding textures of the baseball field with high visual fidelity. The presence of sharp edges and consistent color distribution further underscores the model’s capacity to prioritize semantically critical features during compression. 


\section{Conclusion}
\label{section7}
The integration of SemCom with satellite-ground networks holds significant promise for the development of ComAI. In this paper, we have proposed the S-MDMA model, specifically designed for bandwidth-constrained multi-user satellite-ground communication. This model first performs semantic extraction and merging to identify and integrate similar semantic components. To improve transmission efficiency, a semantic importance sorting strategy based on reconstruction sensitivity has been introduced, allowing for the selective retention of the most critical semantic features. Furthermore, to address the inherent practical challenges in multi-user SemCom, we have introduced orthogonal embedding of semantic features and designed a multi-user reconstruction loss function to mitigate inter-user interference. Extensive experiments on open-source datasets have demonstrated that the proposed S-MDMA model consistently outperforms existing methods and exhibits robust performance under various transmission conditions.

Although the proposed S-MDMA framework is effective, there are still some aspects that need further research. Fisrt, The current design targets a two-user setting. Future work can extend S-MDMA to support scalable multi-user access ($M > 2$) by incorporating hierarchical fusion and adaptive user grouping. Additionally, dynamic scheduling strategies based on channel quality, task priority, and semantic relevance can be explored to enhance fairness and resource utilization under varying user loads. Then, to support diverse satellite-ground applications, future research can integrate cross-modal semantics, such as image and text fusion, into the transmission process. A unified embedding space and multi-modal encoder can achieve effective joint representation and compression of different types of data, thereby improving task adaptability and semantic completeness. Third, actual deployment in satellite-ground systems remains a crucial step. Future research can evaluate the performance of S-MDMA under realistic orbital environments, atmospheric interference, and hardware limitations. 


\bibliography{ref} 

\bibliographystyle{IEEEtran}


 




\vfill

\end{document}